\documentclass[structabstract]{aa}  
\usepackage{graphicx,natbib}

\usepackage{txfonts}
\begin{document}
   \title{The Milky Way's external disc constrained by 2MASS star counts}

%   \subtitle{I. Overviewing the $\kappa$-mechanism}

   \author{C. Reyl\'e
          \inst{1}
          \and
          D. J. Marshall\inst{2}
          \and
          A.C. Robin\inst{1}
          \and
          M. Schultheis\inst{1}
          %\fnmsep\thanks{Just to show the usage
          %of the elements in the author field}
          }

   \institute{Observatoire de Besan\c{c}on, Institut UTINAM, Universit\'e de Franche-Comt\'e, BP1615, F-25010 Besan\c{c}on cedex, France\\
              \email{celine@obs-besancon.fr}
         \and
             D\'epartement de Physique et Centre de Recherche en Astrophysique de Qu\'ebec, Universit\'e Laval, Qu\'ebec, G1K 7P4, Canada\\
%             \email{c.ptolemy@hipparch.uheaven.space}
%             \thanks{The university of heaven temporarily does not
%                    accept e-mails}
             }

   \date{Received ; accepted }

% \abstract{}{}{}{}{} 
% 5 {} token are mandatory
 
  \abstract
  % context heading (optional)
  % {} leave it empty if necessary  
   {Thanks to recent large scale surveys in the near infrared such as 2MASS, the galactic plane that most suffers from extinction is revealed and its overall structure can be studied.}
  % aims heading (mandatory)
   {This work aims at constraining the structure of the Milky Way external disc as seen in 2MASS data, and in particular the warp.}
  % methods heading (mandatory)
   {We use the Two Micron All Sky Survey (hereafter 2MASS) along with the Stellar Population Synthesis Model of the Galaxy, developed in Besan\c{c}on, to constrain the
external disc parameters such as its scale length, its cutoff radius, and the slope of the warp. In order to properly interpret the observations,
the simulated stars are reddened using a three dimensional extinction map. The shape of the stellar warp is then compared with previous results and with similar structures in gas and dust. 
}
  % results heading (mandatory)
   {We find new constraints on the stellar disc, which is shown to be asymmetrical, similar to observations of HI. The positive longitude side is found to be easily modelled with a S shape warp but with a slope significantly smaller than the slope seen in the HI warp. At negative longitudes, the disc presents peculiarities which are not well reproduced by any simple model. Finally, comparing with the warp seen in the dust, it seems to follow a slope intermediate between the gas and the stars.}
  % conclusions heading (optional), leave it empty if necessary 
   {}


   \maketitle
%
%________________________________________________________________

\section{Introduction}
In recent years wide surveys have been obtained thanks to wide-field mosaic of CCD cameras
and helped by the availability of dedicated telescopes both from the ground and from space. These large data sets benefit Galactic structure studies.

However, interstellar extinction remains a serious obstacle for the observation
of stars in the Milky Way, and for interpreting these observations
in terms of Galactic structure. As the extinction suffered
in the near infrared is roughly an order of magnitude lower than in the visible,
recent infrared surveys provide a suitable tool to probe further into the densest parts of the Galaxy, namely the Galactic plane.

Among them, the 2MASS survey \citep{Skrutskie2006} has provided one of these rich data sets from which substantial results have been obtained. The homogeneity of these data sets is a great help for avoiding systematic bias and bad links from one set to another, which have created troubles in data interpretation in the past, disentangling real cosmic substructure from local zero point errors for example. 

The analysis of near infrared data in the whole Galactic plane can bring important clues regarding the stellar populations. In particular, a good knowledge of the disc structure is crucial to characterize the overdensities found close to the Galactic plane, such as the Canis Major overdensity. This structure located at $l=240^\circ$, $b=-8^\circ$ was revealed by \cite{Martin2004a}. Its nature is still under debate. While several studies based on kinematics, star counts and CMD studies of this
field conclude that it is a dwarf galaxy \citep{Martin2004b,Martin2005,Martinez2005,Butler2007,DeJong2007}, other authors tested several warp models and claimed that the warp is still a
possible explanation of the Canis Major overdensity \citep{Momany2004,Momany2006,Lopez2006}. It shows how important it is to have a good description of the Galactic disc structure. 

Warps are a common feature in external galaxies \citep{Bosma1991}. Radio observations revealed that the Milky Way itself hosts a warp and flared disc. The warp and flares have been observed in HI data \citep{Henderson1982, Burton1986, Burton1988, Diplas1991,Nakanishi2003,Levine2006}, in molecular clouds \citep{Grabelsky1987,Wouterlout1990,May1997}, from OB stars \citep{Miyamoto1988,Reed1996} and finally in COBE data \citep{Porcel1997,Freudenreich1998,Drimmel2001}. The warp and flare in the Galaxy have also been detected in the interstellar dust \citep{Freudenreich1994,Marshall2006}.
\cite{Derriere2001} have analysed DENIS \citep{Epchtein1999} star counts in several fields in the Galactic plane and showed evidence of a warp and a flare in the stellar disc. More recently \cite{Lopez2002} modelled a stellar warp from the analysis of the stellar populations in the 2MASS survey.

Warps may have originated from interactions between the disc and (i) the dark halo (if angular momenta are not aligned), (ii) nearby satellite galaxies, such as the Sagittarius dwarf or the Magellanic clouds, 
(iii) infalling intergalactic gas. From the analysis of their angular momenta,
 \cite{Bailin2003} argues that the Milky Way warp and the Sagittarius Dwarf 
Galaxy may be coupled. \cite{Garcia-Ruiz2002} used N-body simulations to explore the possibility that the warp has been created by tidal forces caused by the Magellanic Clouds. They find that neither orientation nor amplitude
of the warp can be reproduced this way. 
On the contrary, \cite{Weinberg2006} using perturbation theory showed that a Magellanic Cloud origin for the warp of the Milky Way can explain most quantitative features of the outer HI layer.
Alternatively \cite{Lopez2002} propose that the warp is due to intergalactic accretion flows onto the Milky Way disc.

Here we report on an the analysis of the external stellar disc of the Galaxy, and particularly the warp, as seen from near infrared star counts using a Galaxy model. We overview the stellar population synthesis approach and the basic inputs of the model, in particular for the external disc (Sect.~\ref{mod}). In Sect.~\ref{ext} we briefly describe the determination of a 3D map of interstellar extinction and show a first comparison between the Galaxy model and the 2MASS survey in the Galactic plane.
We then describe (sect.~\ref{result}) constraints obtained from the comparisons of model predictions using different disc models with 2MASS data and compare this with results obtained from other studies. In sect.~\ref{comp} we provisionally conclude on the comparison between the gaseous warp and the stellar warp and about the origin of this warp.

\section{The Galaxy model}
\label{mod}

\subsection{Stellar population synthesis approach}

In order to analyse the large data set provided by 2MASS, we use a stellar population synthesis approach (the Galaxy model, \cite{Robin2003}). This model aims at assembling current scenarii of galaxy formation and evolution, 
theories of stellar formation and evolution, models of stellar atmospheres and
dynamical constraints, in order to make a consistent picture
explaining currently available observations. When scenarii of Galaxy formation and evolution are inferred from suitable constraints, they may be tested using population synthesis models whose predictions can be directly compared with observations. This synthetic approach ensures that biases have been correctly taken into account and that the scenario is compatible with many kinds of constraint.

The validity of any Galactic model is always questionable, as it describes a smooth Galaxy, while inhomogeneities exist, either in the disc or the halo. The issue is not to make a perfect model that reproduces the known Galaxy at any scale. Rather one aims to produce a useful tool to 
compute the probable stellar content of large data sets and therefore to test the 
usefulness of such data to answer a given question relative to Galactic structure and evolution. 

The model we use is the Galaxy model developed in Besan\c{c}on. Its originality, as compared to a few other population synthesis models presently available for the Galaxy, is the dynamical 
self-consistency. The Boltzmann equation allows the scale height
of an isothermal and relaxed population to be constrained by its velocity
dispersion and the 
Galactic potential \citep{Bienayme1987a}. 
The use of this dynamical constraint avoids a set of 
free parameters quite difficult to determine: the scale height
of the thin disc at different ages. 

\subsection{Basic scheme}
The model assumes that stars belong to four main 
populations: the thin disc, the thick disc, the 
stellar halo (or spheroid), and the outer bulge.
The modelling of each population is based on a set of evolutionary tracks,
assumptions on density distributions, constrained either by dynamical
considerations or by empirical data, and guided by a scenario of
formation and evolution, that is to say assumptions on 
the initial mass function (IMF) and the star formation rate (SFR) for each population. 
More detailed descriptions on these constraints can be found in \cite{Haywood1997} for the thin disc, \cite{Reyle2001} for the thick disc, \cite{Robin2000} for the spheroid, and \cite{Picaud2004} for the outer bulge. 
Recently, the Hipparcos mission and large scale surveys in the optical
and the near-infrared have led to new physical constraints improving our
knowledge of the overall structure and evolution of the Galaxy. These
new constraints are included in the version of the model described in \cite{Robin2003}.

\subsection{The external disc model}
\textbf{The disc is assumed to be axisymmetric and spiral arms are not modelled. The thin disc is divided into 7 age
components. The distribution of each disc component is described by an axisymmetric ellipsoid with an axis ratio depending on the age, using an Einasto law}.
The disc scale length, $h_R=$2530 pc, is derived  from the fitting of the old disc parameters on DENIS star counts towards the Galactic bulge direction \citep{Picaud2004}. Several studies have shown that the edge of the disc is detected at a galactocentric distance of about 14 kpc \citep{Robin1992,Ruphy1996}.
The disc is also warped and flared \textbf{and all stellar components in the disc have the same warp and flare}. 

As in \cite{Gyuk1999}, we model the flaring by increasing the scale heights by a factor $k_\mathrm{flare}$, beyond a galactocentric radius $R_\mathrm{flare}$:
$$k_\mathrm{flare}(R)=1+\gamma_\mathrm{flare}(R-R_\mathrm{flare})$$
The amplitude is taken from \citet{Gyuk1999} ($\gamma_\mathrm{flare}=5.4 \times 10^{-5}$ pc$^{-1}$)
and the radius $R_\mathrm{flare}$ at which the disc starts flaring from \cite{Derriere2001}.They found using DENIS data that the minimum radius of the flare might depend on the longitude considered and determined an average value of $R_\mathrm{flare}=9.5$~kpc.

We first model the warp as a S shape warp, with a tilted ring model following \cite{Porcel1997}. 
The height $z_\mathrm{warp}$ of the warp over the plane b=0$^\circ$ as a function of galactocentric coordinates is computed using:
$$z_\mathrm{warp}(R) = \gamma_\mathrm{warp} \times (R-R_\mathrm{warp}) \times \sin(\phi-\phi_\mathrm{warp})$$

Following \cite{Burton1988}, 
we assume, as a starting point, that the Sun lies on the line of nodes of the warp 
($\phi_\mathrm{warp} = 0^\circ$). Most of warp studies indicate values of this angle close to this one within an uncertainty of a few degrees. 
We use $R_\mathrm{warp}=8.4$~kpc indicated as the best value for the starting galactocentric radius of the warp by \cite{Derriere2001}. This value is close to the galactocentric distance of the Sun as adopted in the Galaxy model.
We adopt, as an initial guess, the value of the displacement of the mid plane $\gamma_\mathrm{warp}=0.18$ from \cite{Gyuk1999}. It gives positions of the mid-plane similar to the \cite{Lopez2002} and \cite{Drimmel2001} stellar warp models, with a maximum height of 630 pc at 12 kpc from the Galactic Centre. 

\section{Interstellar extinction}
\label{ext}

The 2MASS survey allows us to study large scale structure in the Galaxy by comparing with simulations of the Galaxy model, particularly in the Galactic plane where extinction is quite high. 
The comparison is realistic only if the extinction is known. Without a good estimate of the extinction and of its distance it is nearly impossible to understand the structure in the thin disc. Hence it is imperative to start by constructing a three dimensional extinction map of the Galaxy.

Extinction is so clumpy in the Galactic plane that it determines for a great part the number density of stars, more than any other large scale stellar structure. Thus, it is possible to extract information about the distribution of the extinction from photometry and star counts. \cite{Marshall2006} have shown that the 3D extinction distribution can be inferred from the colour distributions in the 2MASS survey. Using stellar colours in $J-K_S$ as extinction indicators and assuming that all of the model prediction deviations on small scales from observed colours arises from the variation of extinction along the line of sight, they built a 3D extinction map of the galactic plane. The final resolution in longitude and latitude is 15 arcmin and the resolution in distance varies between 100 pc to 1 kpc, depending on stellar density and on the dust distribution along the line of sight. The resulting 3D extinction map provides an accurate description of the large scale structure of the disc of dust. These maps clearly show the dust warp in the external disc (Fig.~1). Note that the warp found for the dust differs from the stellar warp model.

\begin{figure*}[ht]
\label{warpdust}
\begin{center}
%\hbox{\hspace{0cm}
\includegraphics[width=\columnwidth]{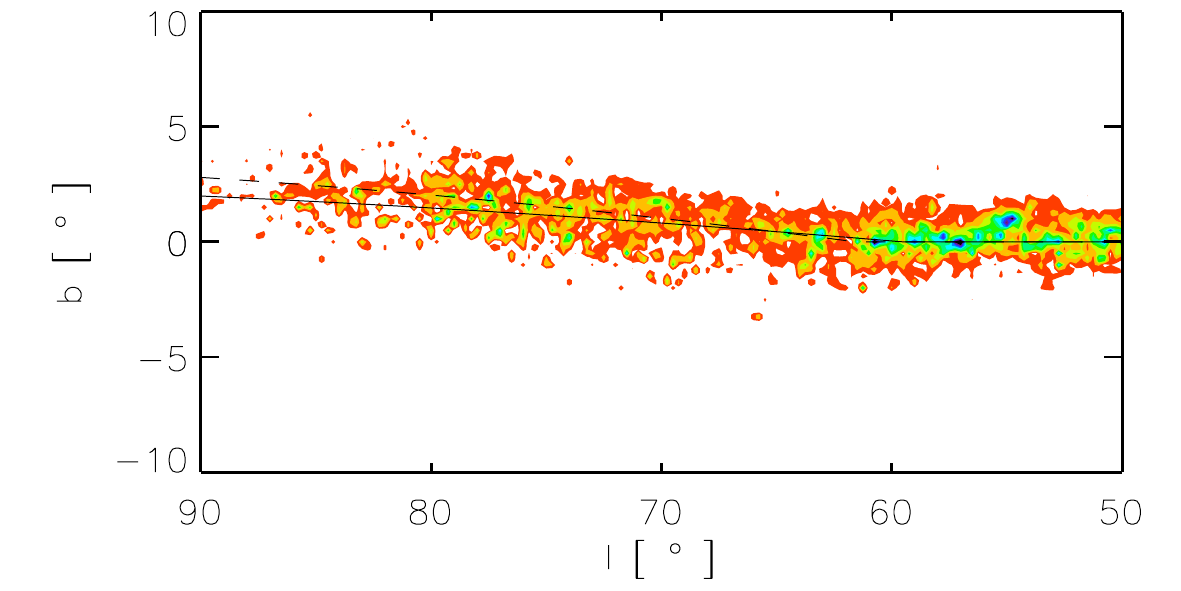}
\includegraphics[width=\columnwidth]{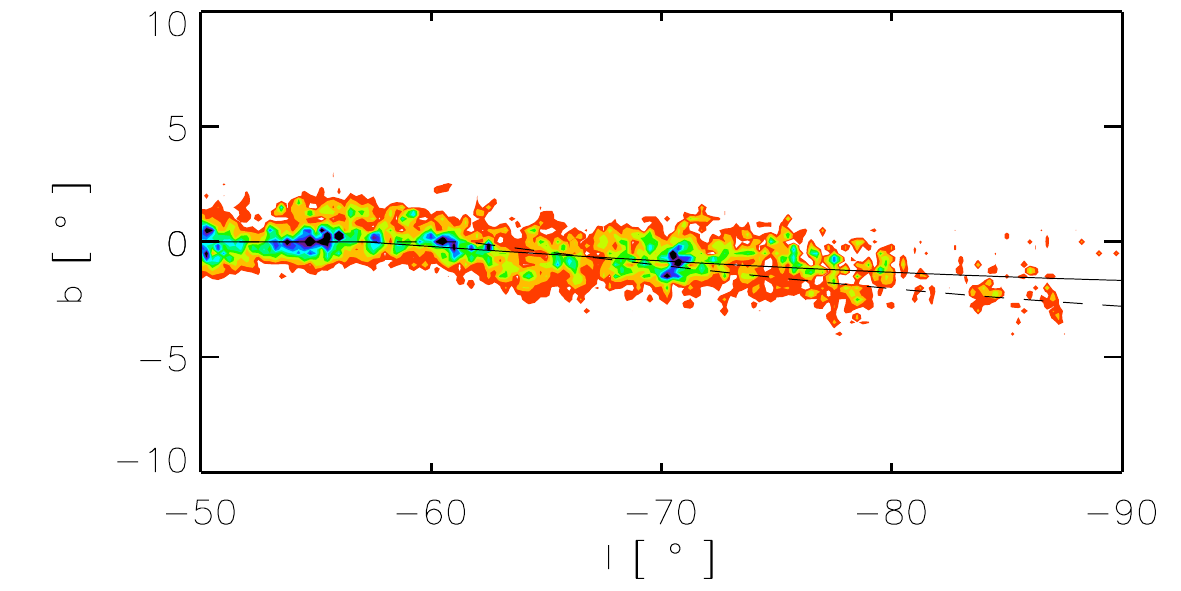}
%}
\caption{Warp in the dust as seen from \cite{Marshall2006} 3D extinction map. The plot shows the density of absorbing material at 8kpc from the sun, as a function of longitude and latitude. The dashed line is the position of the mid-plane expected from a 0.18 slope warp model, and the solid line is the position from the fitted slope of 0.14 at positive longitudes and 0.11 at negative longitudes.}
\end{center}
\end{figure*}

\begin{figure*}[h]
\centering
\includegraphics[width=\textwidth,angle=0]{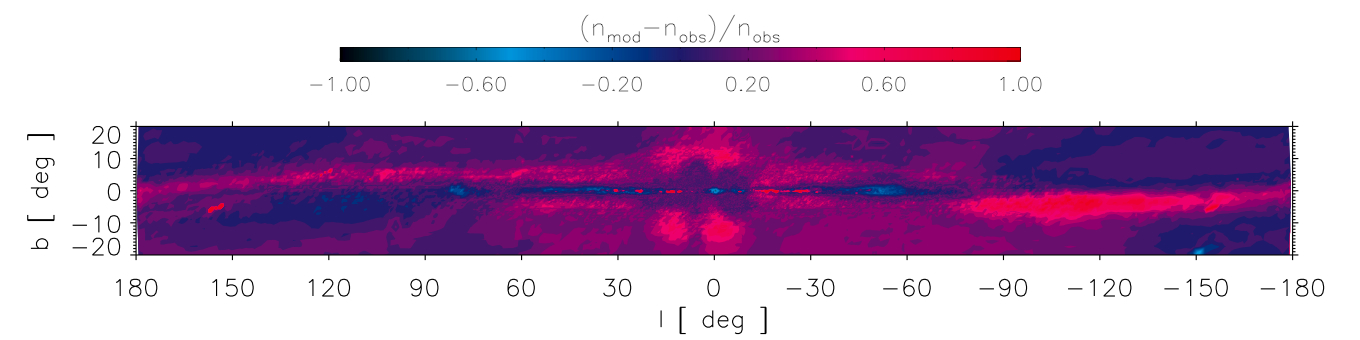}
\caption{Comparison between the number of stars from the Galaxy model $n_\mathrm{mod}$ and 2MASS 
data $n_\mathrm{obs}$. The counts are given in a colour table in number of stars per square degree to 
magnitude $K_S=12$. The relative difference is given as $(n_\mathrm{mod}-n_\mathrm{obs})/n_\mathrm{obs}$. 
It varies from a factor of two excess in modelled star counts to a factor of two deficiency in modelled 
star counts. Regions where the Galaxy model overestimates or underestimates de density of stars clearly appear.}
\label{fig1}
\end{figure*}

After applying the extinction to the simulated stars, it is possible to perform a realistic comparison of the number $n_\mathrm{mod}$ of stars from the Galactic model with the number $n_\mathrm{obs}$ of observed stars in the 2MASS data (Fig.~\ref{fig1}). The comparison points to significant discrepancies.
Thus it appears that several galactic components are not properly characterized in the
Galaxy model: (i) the bulge extends too far from the plane, (ii)  the disc in the inner part of the Galaxy ($|$l$|<$75$^\circ$) has a scale height slightly too large, (iii) the warp of the external disc is too strong, and 
(iv) the modelled warp is symmetrical whereas the observations point to an asymmetrical warp.

In this work we concentrate on the external disc ($|$l$|>$60$^\circ$) and propose to derive a better
parametrization in three steps: (i) simulations of the Galaxy are obtained using different parameters, (ii) \textbf {as the extinction found depends on the parameters in the Galaxy model (the sensitivity of the extinction model to the Galaxy model is discussed in \cite{Marshall2006}), the extinction is  derived for each set of Galactic parameters} using the \cite{Marshall2006} method and applied to the simulated stars, (iii) number counts of simulated and observed stars are statistically compared.

\section{Results}
\label{result}

As seen in Fig.~\ref{fig1}, the assumed slope of the warp is not suitable for correctly reproducing the data. The slope value is too high and the simulated warp deviates too far from the $b=0^\circ$ plane.
Too many stars are predicted below the galactic plane at negative longitudes and above the plane at positive longitudes. The overestimate is higher for $l<0^\circ$, showing that the $\textbf{stellar}$ warp is asymmetric.
\textbf{This can be either an apparent asymmetry if the Sun does not lie on the line of nodes, or a real asymmetry in the disc. However the effect of the position of line of nodes does not solve the problem, as explained further below. Another possibility is that the asymmetry is due to the distribution of the dust with respect to the stars. In our approach we assumed as a first approximation that a symmetric warp exists in the stars only. The extinction we derive does not suppose any geometry for the large scale distribution of the dust - it simply finds the extinction necessary to correct the simulated stellar colours to fit the observed ones. The dust warp then appears naturally in the extinction distribution and this dust warp is not symmetrical.
Assuming this dust warp, we then tentatively ajust the star counts on both sides of the galaxy,
in order to check the values of the northern and the southern warps independently.
As such, it is not possible for us to create a stellar overdensity in the model by a misalignment of the stellar and dusty warps.}

\begin{figure*}[ht]
\centering
\includegraphics[height=9.5cm,angle=0]{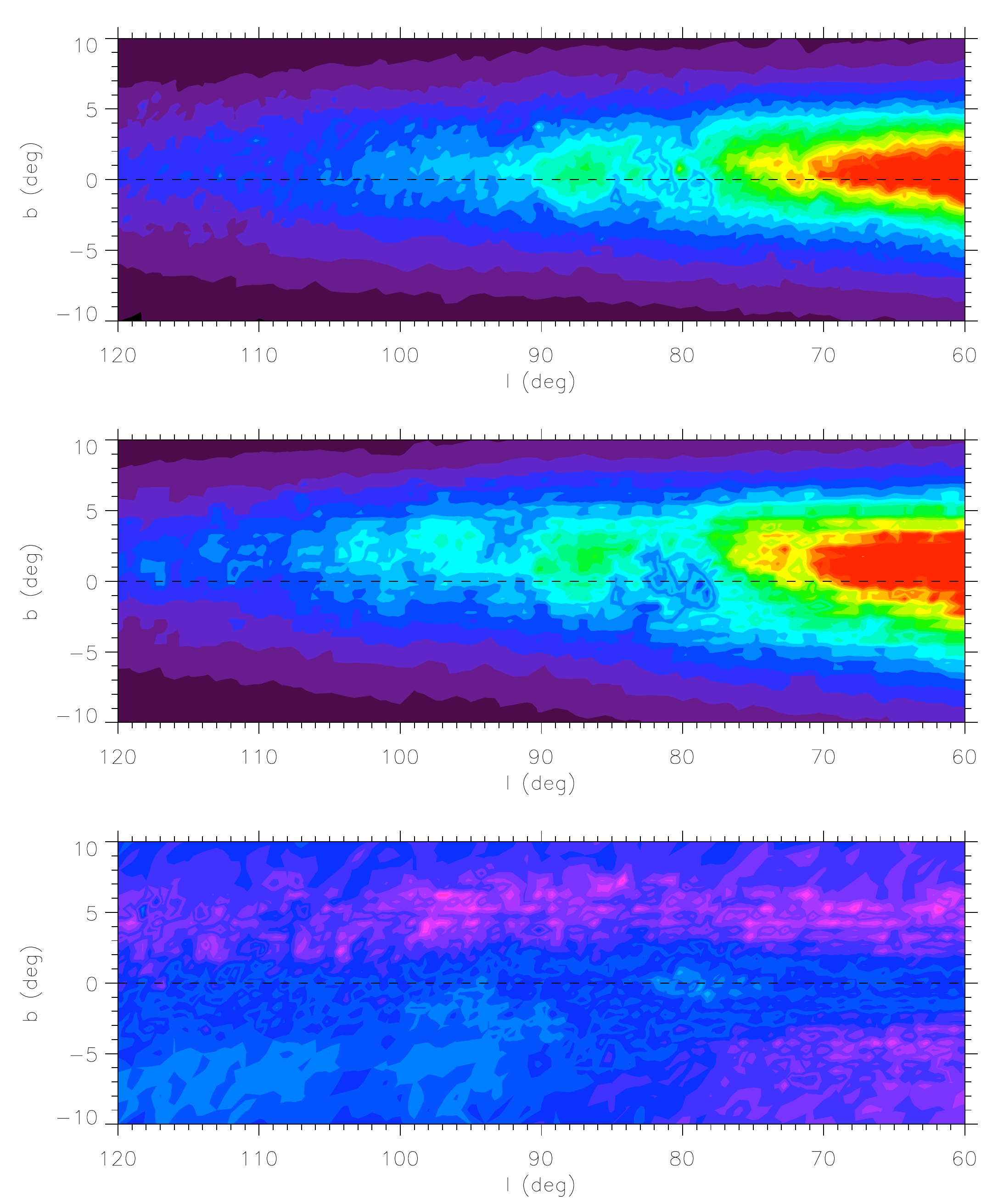}
\includegraphics[height=9.5cm,angle=0]{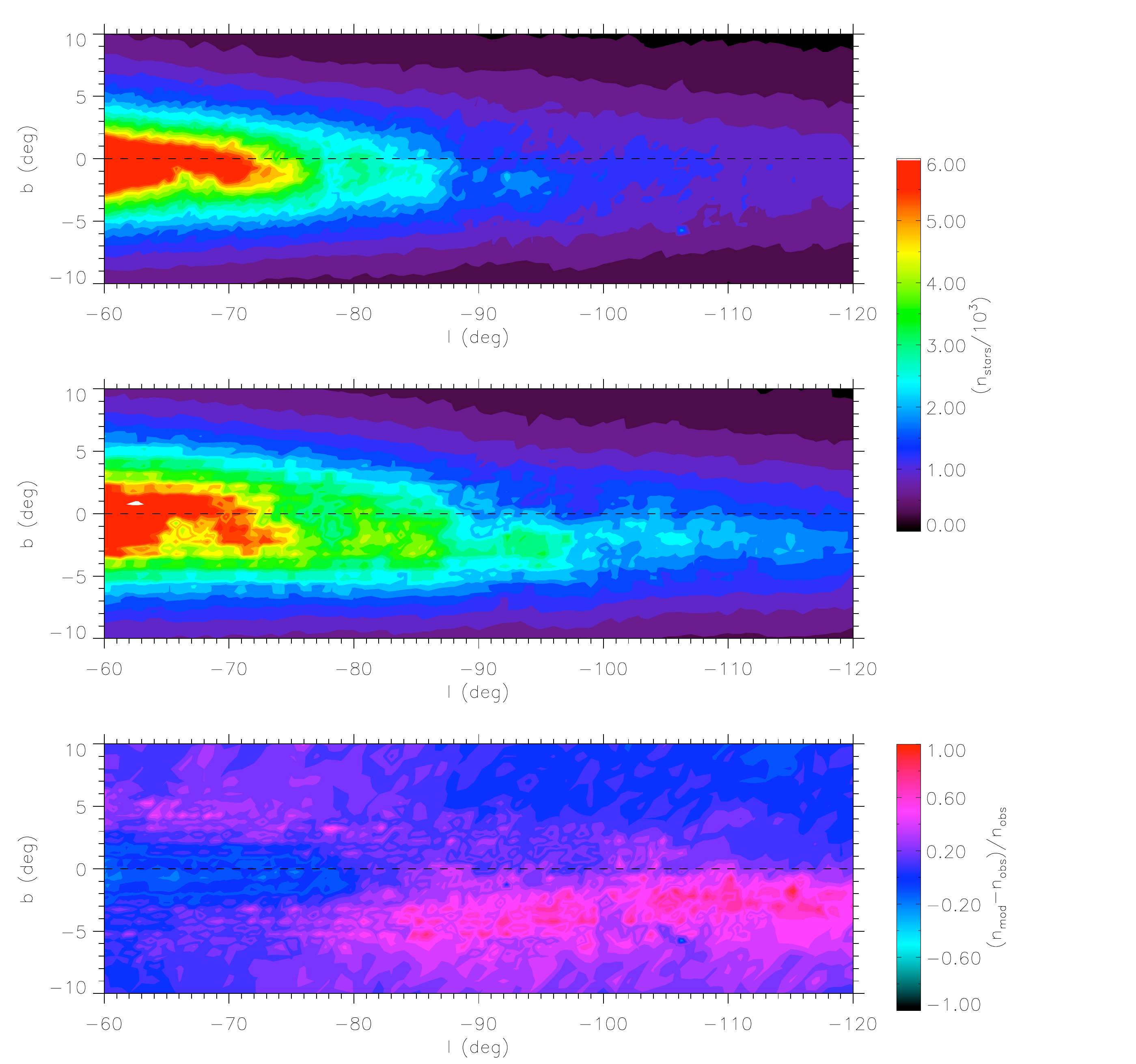}
\caption{2MASS star counts (top), modelled ones (middle), and relative difference between the two (bottom) with $\gamma_\mathrm{warp}$=0.18 and $h_R$=2530 pc. Dashed line indicates the $b=0^\circ$ plane. On the left the Northern warp, on the right the Southern warp.}
\label{fig2}
\end{figure*}

Fig.~\ref{fig2} (top panel) shows 2MASS star counts in the longitude ranges 120$^\circ$ to 60$^\circ$ and $-60^\circ$ to $-120^\circ$.
They are compared with simulated star counts obtained with the simulations from the
Galaxy model (middle panel). The difference is shown in the bottom panel. In addition to the
overestimate of the warp slope, it appears that the modelled density of stars is too large at  $|l| \gtrsim 80^\circ$
probably due to the value of the scale length $h_R$ being too high.

\begin{figure*}[ht]
\centering
\includegraphics[height=9.5cm,angle=0]{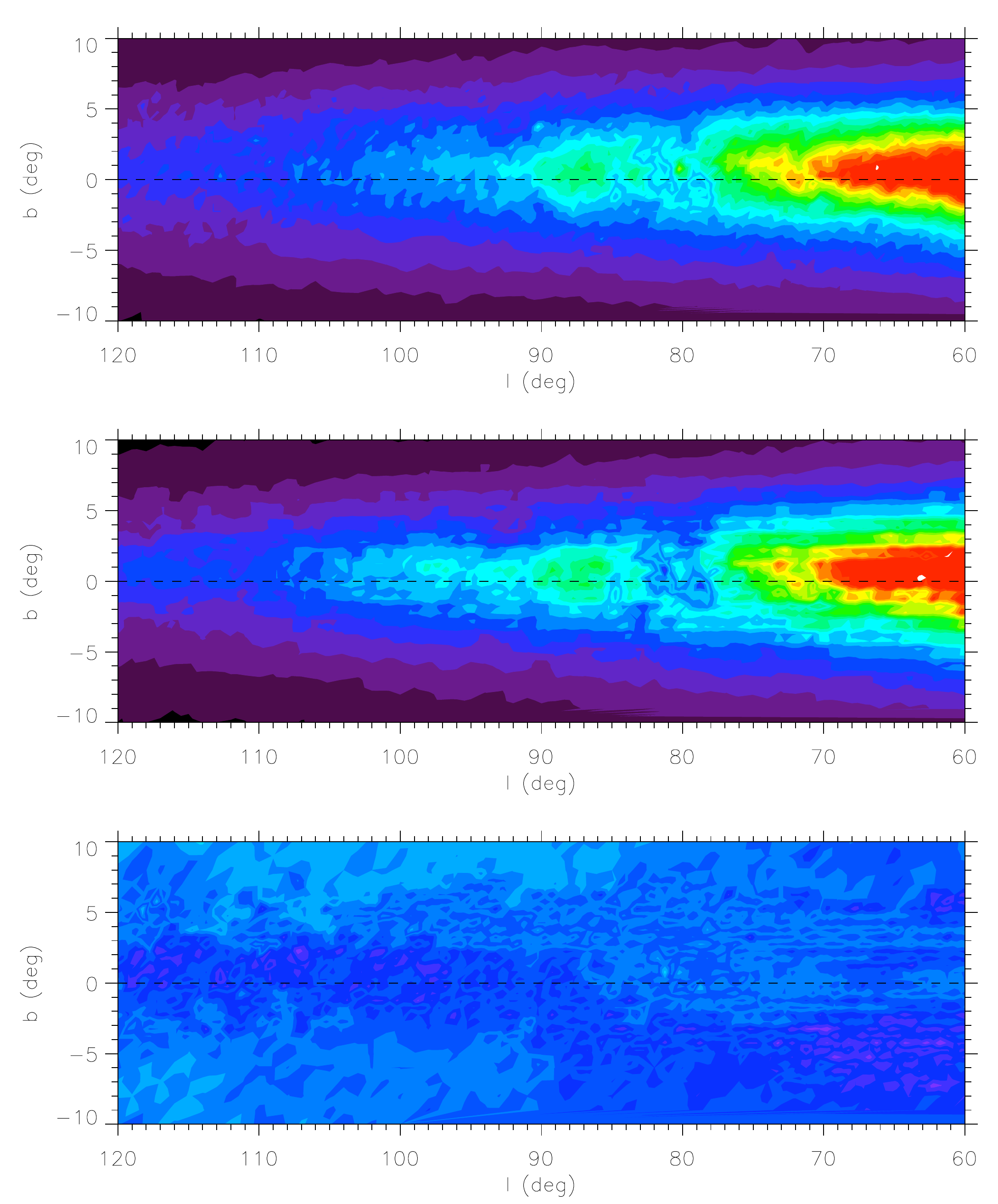}
\includegraphics[height=9.5cm,angle=0]{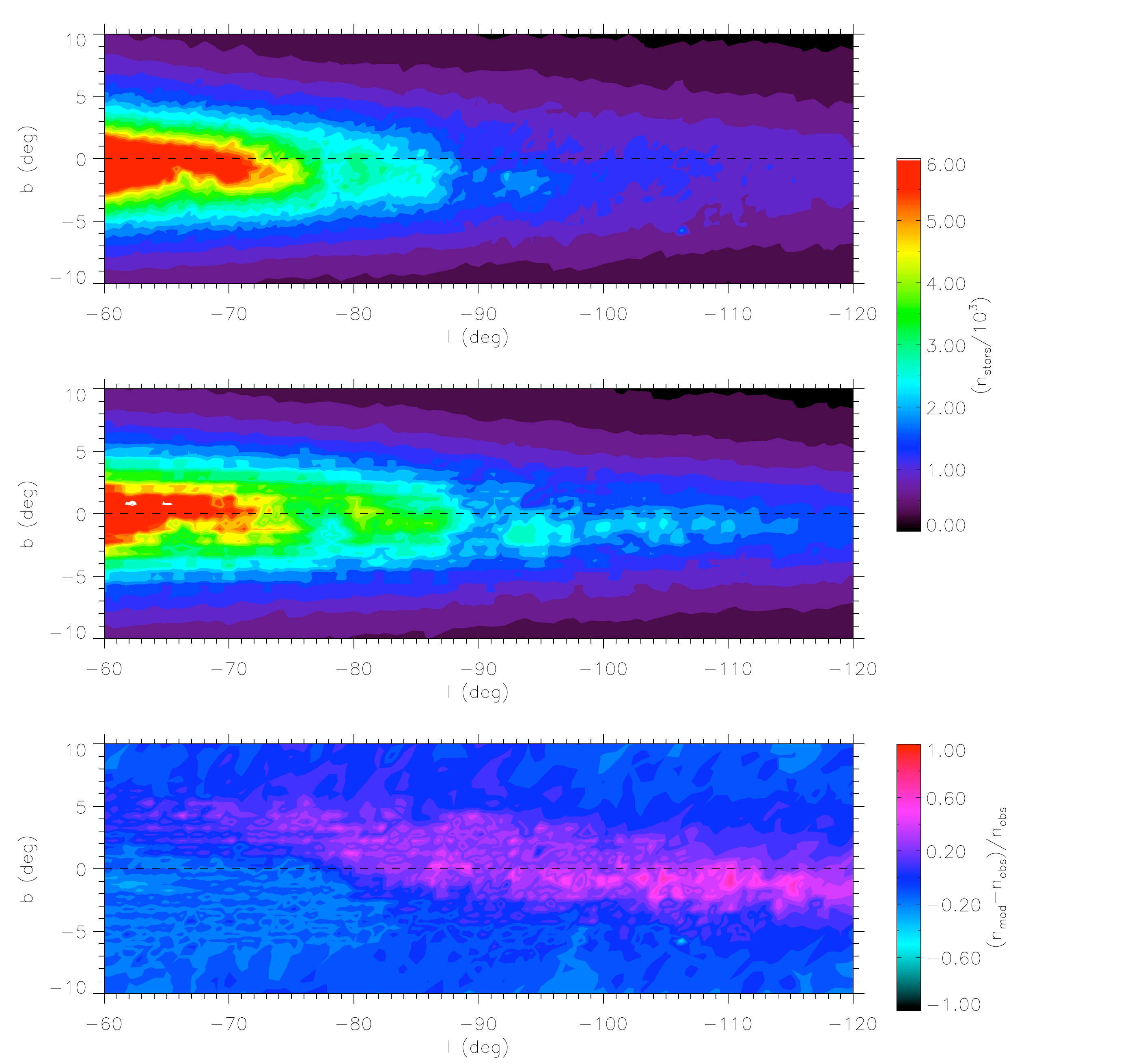}
\caption{Same as Fig.~\ref{fig2} with $\gamma_\mathrm{warp}$=0.09 and $h_R$=2200 pc. The model which reasonnably fits the positive longitude side significantly differs from the data at the negative longitudes, showing a clear disymmetry between both sides of the Galactic disc.}
\label{fig3}
\end{figure*}

A new simulation with the smaller values $\gamma_\mathrm{warp}=0.09$ and $h_R=$ 2200 pc is shown in Fig~\ref{fig3}.
It substantially improves the agreement between model and data at positive longitudes. The value $h_R=$2200 pc is in agreement with the one determined by \cite{Robin1992} from star counts towards the anticenter.

However the comparison is still unsatisfactory at negative longitudes. 
There remains a significant discrepancy, which can not be improved by varying the assumed slope, the disc scale length or the sun-centre distance (assumed to be here 7.9 kpc). Further testing by varying the line of nodes parameter $\phi_\mathrm{warp}$ (from $-30^\circ$ to $30^\circ$) shows that it cannot account for this disc asymmetry either (for the extreme values, the fit is even degraded).

At negative longitude, this model parametrization (S shapoe warp) is clearly unsatisfactory. In an attempt to improve the fit at negative longitudes, we investigate other warp models, in particular non-axisymmetric ones. It is well known that the HI warp is 
asymmetric: the gas warp bends
back to the Galactic plane in the southern hemisphere at R$>$15 kpc (Burton 1988). The stellar warp may also follow this feature. 

\cite{Levine2006} studied in detail the warp structure from HI surveys and
found that the galactic HI warp is well described by a vertical offset plus two Fourier modes of frequency $m$=1 and 2, the $m$=2 mode accounting for the asymmetry. 
The height of the mid plane from the $b=0^\circ$ plane takes the following form as a function of the azimuth $\phi$ :

$$z(\phi)=z_0+z_1 sin(\phi-\phi_1) + z_2 sin(2\phi-\phi_2)$$

where  $z_0$, $z_1$ and $z_2$ are the heights of the mid plane, quadratic functions of the 
galactocentric radius, $\phi_1$ and $\phi_2$ are the node lines of the Fourier modes 1 and 2, all of which are adjusted parameters.

With the best fit parameters from \cite{Levine2006} the m=1 mode dominates up to about R$<$15 kpc. Closer to the Galactic centre the other modes are unimportant. The m=2 mode is significant at R$>$20 kpc. This means that the detection of these modes might not be easy with the 2MASS survey if the stellar warp follows the gas warp. Red clump giants at $K_S=14.3$ (the limiting magnitude of 2MASS survey) reach distances of only about 10 kpc from the sun.

In order to test whether the stellar populations might also follow this gaseous warp, we tentatively modelled the stellar warp with the \cite{Levine2006} model. If we adopt their parameters, the simulated counts do not give an acceptable fit. However if we adopt a closer value for the starting galactocentric radius of the m=2 mode, 12 kpc instead of 15 kpc, and reduced slopes, the fit at negative longitude is slightly better and still good at positive longitude, although not as good as the tilted ring model with 
$\gamma_\mathrm{warp}$=0.09 (Fig.~\ref{fig3}). The result is shown in Fig.~5. The systematic discrepancy remains which is not explained by the two Fourier modes warp model.  
 
\begin{figure*}[ht]
\begin{center}
\includegraphics[height=9.5cm,angle=0,clip=]{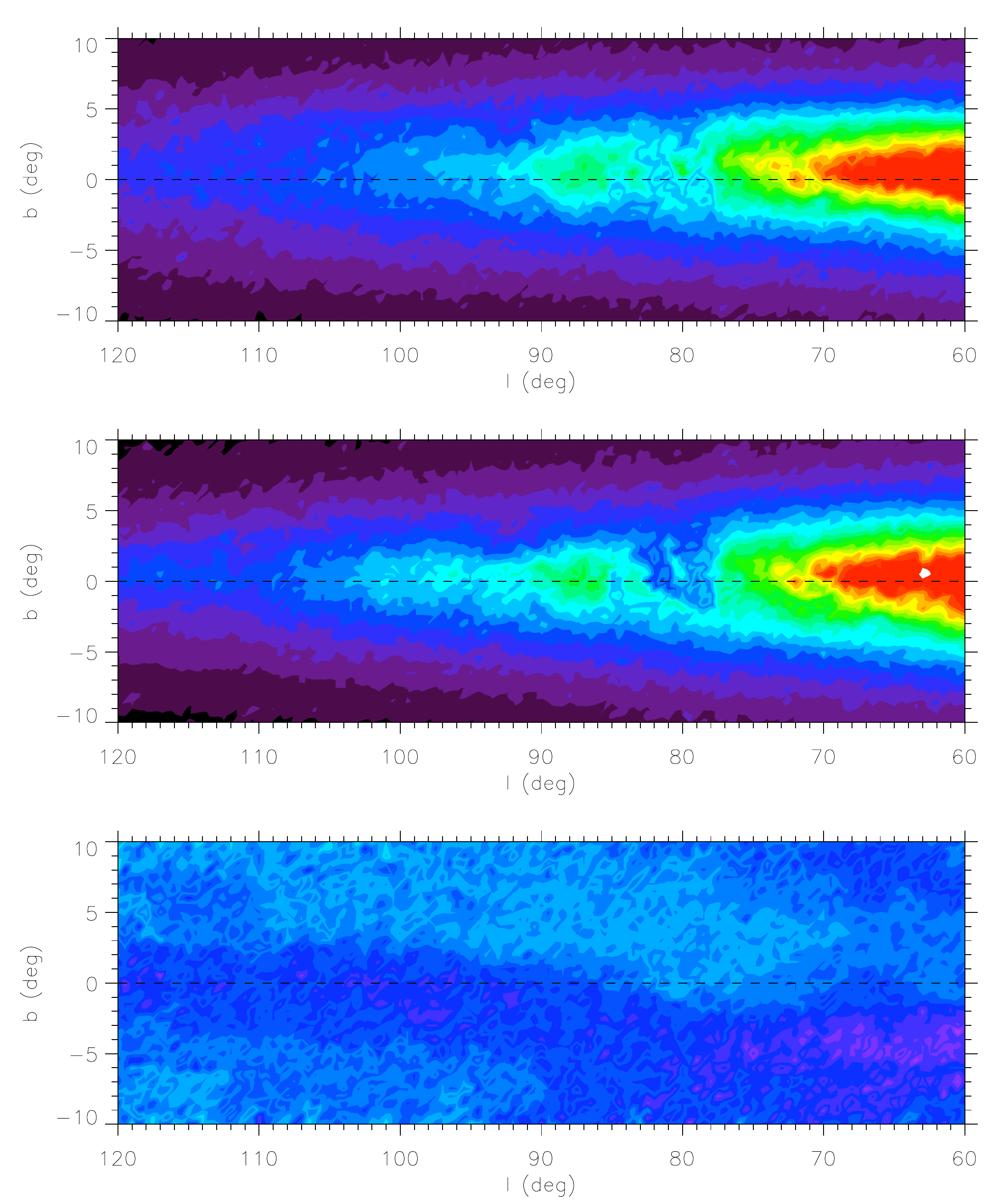}
\includegraphics[height=9.5cm,angle=0,clip=]{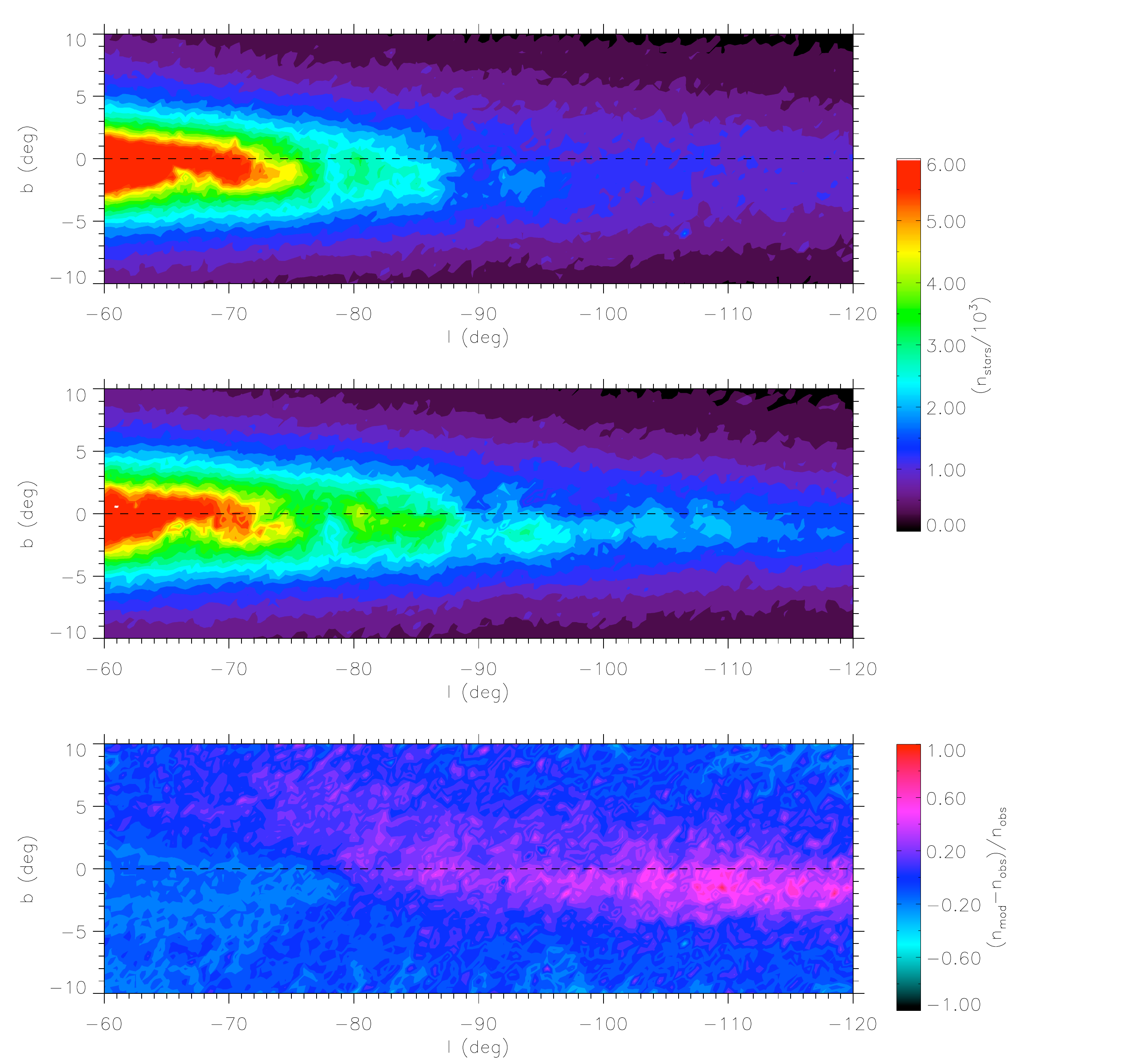}
\caption{Same as Fig.~\ref{fig2} with the two Fourier modes warp model from \cite{Levine2006}. This model does not better fit the stellar data than the simple model and the main discrepancy at negative longitudes is not removed.} 
\end{center}
\label{fig4}
\end{figure*}

We also investigated the possibility of a varying value for the minimum of the flare, such as proposed by \cite{Derriere2001}. However, a lower value $R_\mathrm{flare}=8.5$~kpc does not improve the comparison.

A possible explanation is that the truncation in the stellar disc occurs at shorter distances on this side of the disc, since the standard truncation used on the other side of the disc ($R_c=14$ kpc) creates an overestimation of the star counts at large distances. Indeed, considering a cutoff radius $R_c=12$ kpc
slightly reduces the discrepancies at negative longitudes but still not satisfying (see Fig.~6). Assuming a lower value for $R_c$ causes a deficit of model stars around longitudes $l=70^\circ$.
At positive longitudes, the model with a disc truncation at 12 kpc does not agree as well as a model with $R_c=14$ kpc. This could be an indication of a varying distance of the edge of the disc as a function of the azimuth. The edge of the disc possibly follows an ellipsoidal shape instead of a circular one.

\begin{figure*}[ht]
\begin{center}
\includegraphics[height=9.cm,angle=0,clip=]{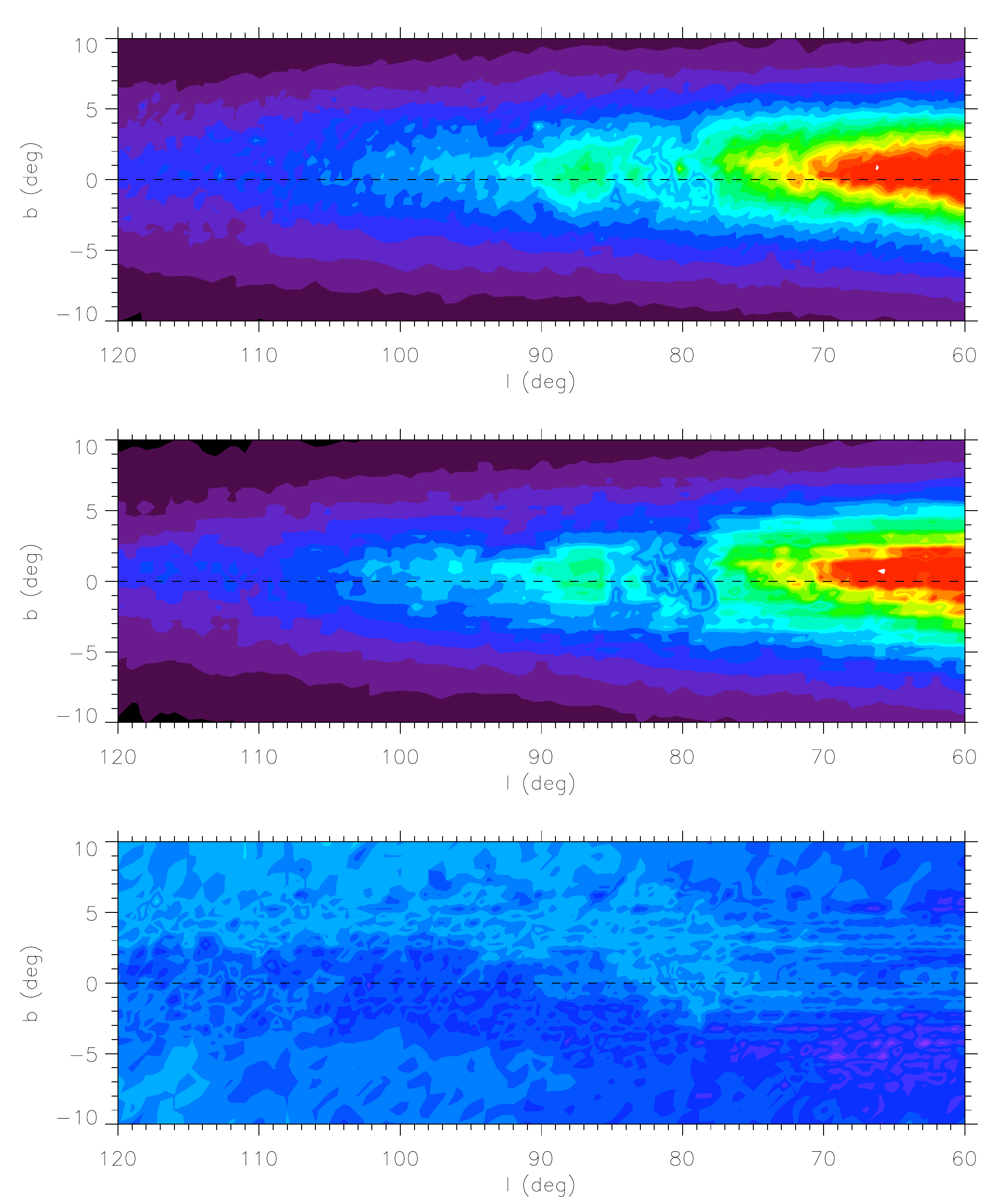}
\includegraphics[height=9.cm,angle=0,clip=]{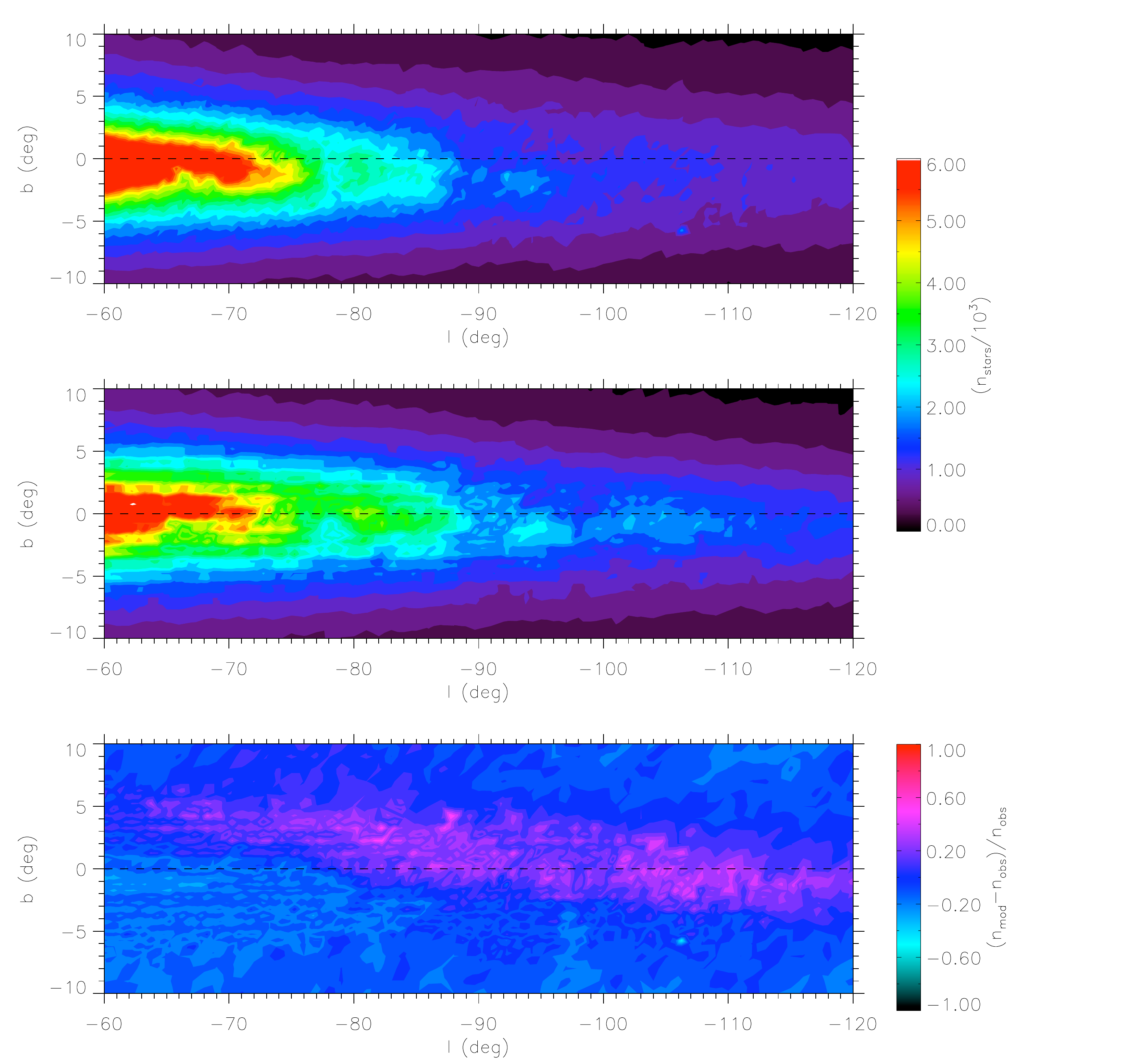}
\caption{Same as Fig.~\ref{fig2} with a truncation of the disc at 12 kpc. The main discrepancy at negative longitudes is slghtly reduced but not removed.} 
\end{center}
\label{figrc}
\end{figure*}

\begin{figure*}[ht]
\centering
\includegraphics[width=\columnwidth,angle=0]{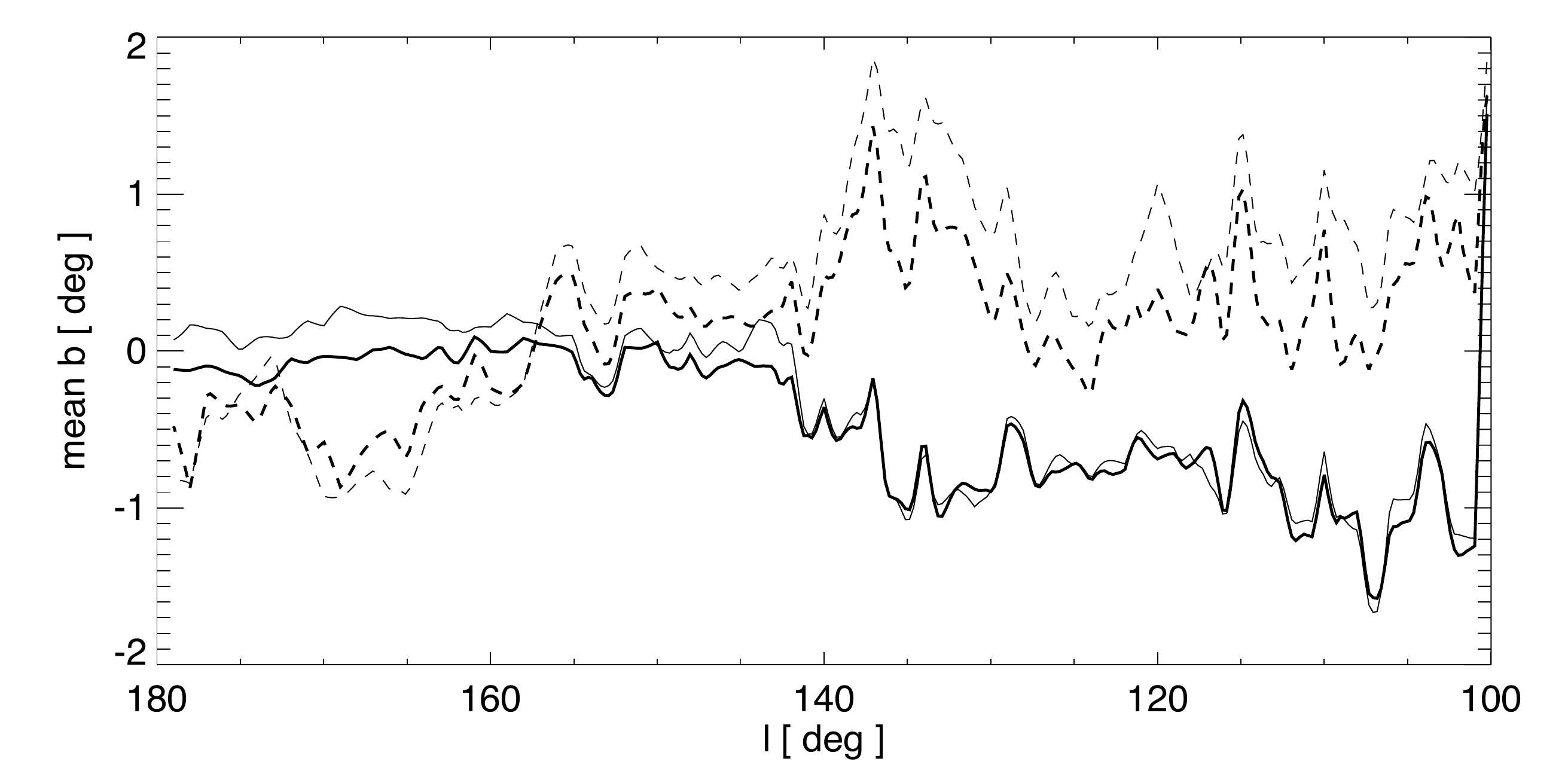}
\includegraphics[width=\columnwidth,angle=0]{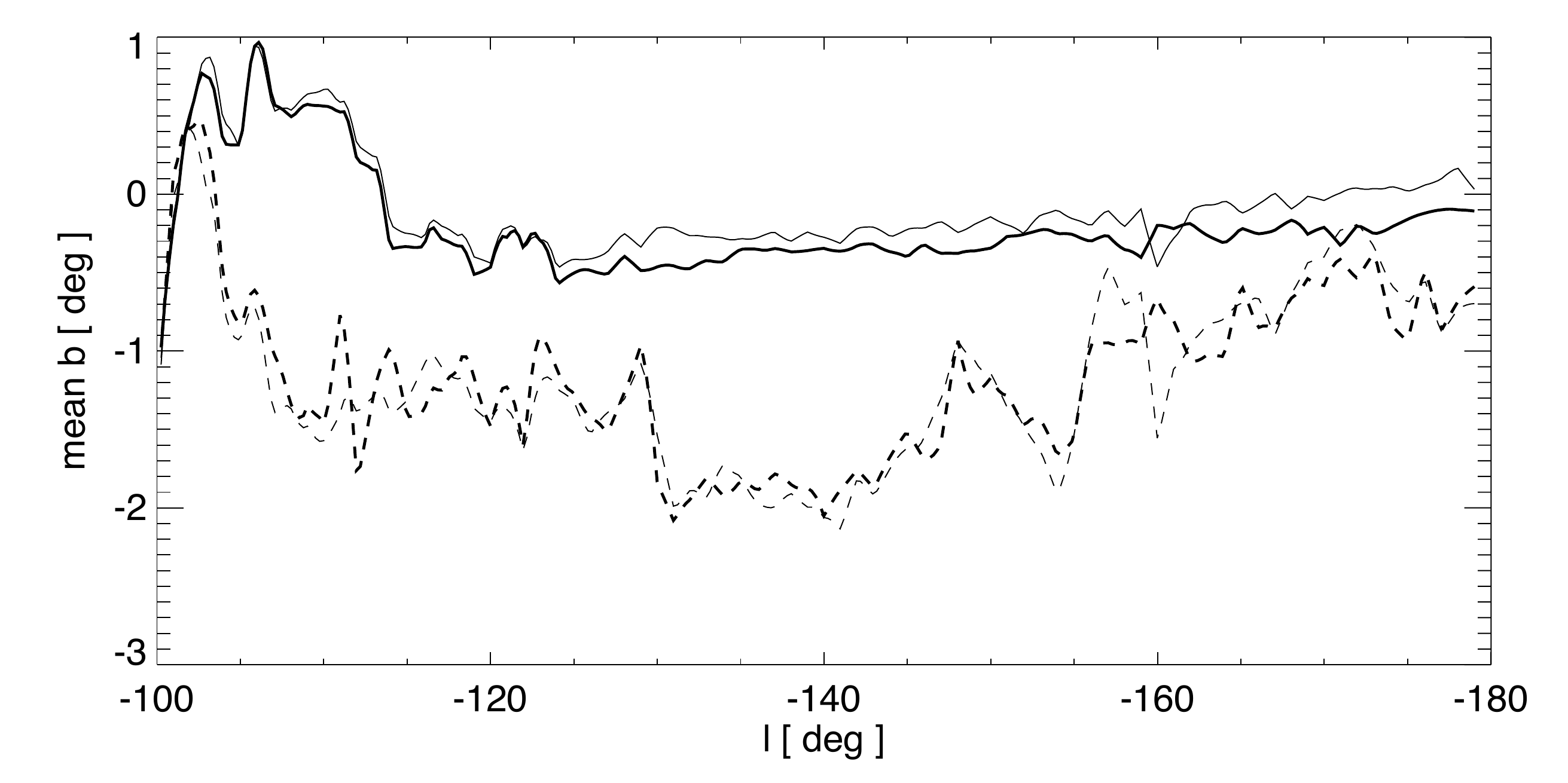}
\caption{Mid plane of data (thin lines) and model (thick lines) for all stars (solid lines)
and giants only (dotted lines). The simulations are obtained with  $\gamma_\mathrm{warp}$=0.09 and $h_R$=2200 pc.}
\label{fig5}
\end{figure*}

The mid plane of stars as a function of longitude is given by: $$\bar{b}(l)=\frac{\sum_b b.\rho(l,b)}{\sum_b \rho(l,b)}$$ where $\rho(l,b)$ is the density of stars.
Fig.~\ref{fig5} shows the mid plane for the data (solid thin line) and the simulations (solid thick line) obtained with $\gamma_\mathrm{warp}$ = 0.09, $h_R$ = 2200 pc. As expected from Fig.~\ref{fig3}, the mid plane of simulated stars is in better agreement with
that of data at positive longitudes. However, the mid plane of simulated stars
is not able to fit the observed one over the whole longitude range. 
We also computed the mid plane of giant stars only (dotted line). \textbf{The selection of giants in 2MASS data is based on their expected colour given the derived extinction}. The giants being the
brightest stars probe a more distant region of the disc. For that reason, the mid plane of
giant stars is further away from the $b=0^\circ$ plane.
At negative longitudes, the slope  $\gamma_\mathrm{warp}$ = 0.09 results in a good fit of the mid plane of giant
stars. However, at positive longitudes, this slope seems to be underestimated when
considering giant stars only. This could be an indication of a varying slope of the warp as a
function of galactocentric distance. Such a model with $z_\mathrm{warp}$ varying with galactocentric distance has been proposed by \cite{Lopez2002}, based on the analysis of 2MASS data.
However, the range of distance over which the warp is constrained is rather small, thus the observed  variation in slope is hardly significant.
\textbf{Another interpretation can be that the warp depends on the age. If young stars are born in a gaseous warp originating from gas dynamics, only young stars will follow the warp. However, if the warp is the response to a perturbation by a dwarf galaxy or a misalignement of the dark halo, all components
should be perturbed independently of their age.}

\section{Discussion}
\label{comp}

The warp is a structure that appears in all components of the Milky Way disc: stellar, dust and gas. Hereafter we review recent studies of these different warps and compare them.

\subsection{The stellar warp}
The maximum height of the mid plane at a galactocentric distance of 12 kpc is 315 pc and smaller than the one found in other studies. 

The three-dimensional model for the Milky Way fitted to the far-infrared and near-infrared data from the COBE/DIRBE instrument by \cite{Drimmel2001}
assumes that the displacement of the warp follows a quadratic function of the galactocentric radius, $z_\mathrm{warp}(R)=27.4(R-R_\mathrm{warp})^2\sin(\phi)$, with $R_\mathrm{warp}$ = 7 kpc. The maximum height of the plane at 12 kpc from the Galactic center is 685 pc.

With an elevation of the stellar warp varying as a power law as a function of the galactocentric distance, $z_\mathrm{warp}(R)=1.2\times10^{-3}R(\mathrm{kpc})^{5.25}\sin(\phi+5^\circ)$, \cite{Lopez2002} found a similar value (556 pc), based on the analysis of the old stellar population using data from the 2MASS survey.

\cite{Momany2006} derived the structure of the stellar warp using 2MASS red clump and red giant stars selected at heliocentric distances of 3, 7 and 17 kpc. Their results are in agreement with \cite{Drimmel2001} and \cite{Lopez2002} results for the Southern warp, whereas the Northern warp (at positive longitudes) is less pronounced. Their results are well in agreement with the \cite{Yusifov2004} model based on the analysis of the distribution of pulsars.
They invoked two factors that may contribute to the appearance of an asymmetric warp. First the Sun is not located on the line of nodes (\cite{Yusifov2004} found a rather high value 
$\phi_\mathrm{warp}=15^\circ$). Second the Northern warp is located just behind the Norma-Cygnus arm and possible variations of the extinction may produce an apparent asymmetric warp.

Fig.~8 shows the height of the galactic disc as a function of longitude at an heliocentric distance of 7 kpc (comparable to Fig.~11 in \cite{Momany2006}) from these studies of the stellar warp.
On the contrary, our model of the Northern warp is in agreement with \cite{Momany2006} but our Southern warp is smaller than in all other studies. 

\begin{figure}[h]
 \label{fig7kpc}
\begin{center}
\includegraphics[scale=0.5,bb=131 123 534 505,clip=]{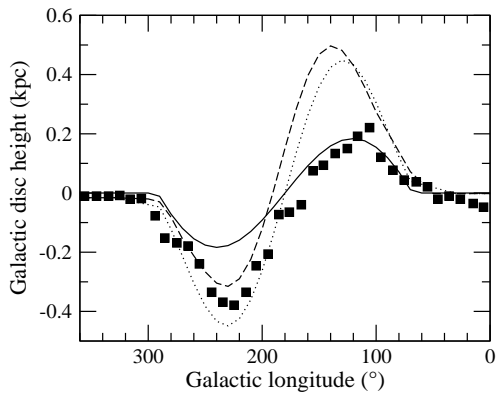}
\caption{Elevation of the disc mid plane as a function of longitude models at an heliocentric distance of 7 kpc from different studies. Solid line: this work $\gamma_\mathrm{warp}=0.09$, dashed line: \cite{Lopez2002}, dotted line: \cite{Drimmel2001}, squares: \cite{Momany2006}.}
\end{center}
\end{figure}
 
\subsection{The dust warp}
\cite{Marshall2006} detected the dust warp when converting the 3D extinction map into interstellar dust volume density. Fig.~1 shows the distribution of dust at 8 kpc from the Sun as a function of longitude. The warp is clearly seen on both sides, but with significant differences. They determined the slope of this dust warp to be 0.14 at positive longitudes and 0.11 at negative ones. They also noticed that the warp seems to start earlier at negative longitudes (at 7.8 kpc rather than 8.9 kpc found on the other side). 

The displacement at a galactocentric distance of 12 kpc is 460 pc, larger than the 315 pc value found for the stellar warp in this work. 
%\cite{Drimmel2001} found that the mid plane of the dust warp is 2 kpc above the $b=0^\circ$ plane at 12 kpc.

\subsection{The gas warp}

\cite{Nakanishi2003} found that the HI disc is warped and asymmetric, with a displacement larger at positive longitudes, in agreement with the dust warp found by \cite{Marshall2006} and the stellar warp in this work. However, the HI warp starts at a galactocentric distance of 12 kpc, which is close to the edge of the stellar disc.

Recently, \cite{Levine2006} used a superposition of Fourier modes to describe the HI disc. The m=1 mode is dominant at distances smaller than 15 kpc. They found that this mode is nearly linear from 10 to 15 kpc, corresponding to a slope of 0.20, much higher than what is found in the dust and stellar warps.

\subsection{Comparison}

These studies reveal a warp present in all components of the disc (stars, dust, gas), all of which are  asymmetric and with a similar line of nodes.
A comparison of the warp elevation as measured in dust, HI gas and stars is given in Fig.~9 together with the galactocentric distance at which the warp starts. In our study, the stellar warp slope is significantly smaller than the HI gas warp, by a factor of about two. There is an indication that the HI gas is the strongest, the dust warp seems a bit smaller and the stellar warp is significantly smaller. However if we limit the investigation to short distances (R$<$12 kpc) the differences between various warp models are less important. It is probable that the difference we find with alternate models \citep{Drimmel2001,Lopez2002,Momany2006} originates from remote counts at galactocentric distances between 12 and 14 kpc. This comparison tends to show that the different components react differently to the forces at the origin of the warp.

\begin{figure}
 \label{figure7}
\begin{center}
\includegraphics[scale=0.5,clip=,bb=100 100 500 500]{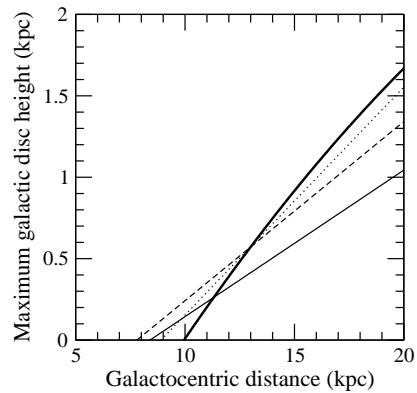}
\caption{Maximum elevation of the disc mid plane from different warp models. Thick solid line: HI \cite{Levine2006} m=1 mode, thin solid line:  stellar warp (this paper), dotted line: \cite{Marshall2006} dust model at positive longitudes, dashed line: \cite{Marshall2006} dust warp at negative longitudes.}
\end{center}
\end{figure}

In external galaxies, \cite{vanderkruit2007} noted that stellar discs look flatter than gas layers. This is understandable in a scheme where the HI warps start close to the truncation radius (truncation seen in the exponential distribution of stars which may be due to a threshold effect in the star formation efficiency). In this case the stellar warp is only visible in a small range of galactocentric radii. \cite{Robin1992} measured the truncation radius in our galaxy to be at about 14 kpc. This is a bit larger than where the HI warp starts, apparently at 10 to 12 kpc, and even larger than where the stellar warp starts (at $\sim$ 8 kpc galactocentric radius). Hence the stellar warp is expected to be detected only between 8 and 14 kpc, at most.

\cite{vanderkruit2007} suggests a scenario where the inner disc formed initially in a rigid and flat structure, while the HI warp resulted from later infall of gas with different orientations of angular momentum (in the case of a dark matter halo not aligned with the one of the inner Galaxy). This on-going infall could be identified with the HI flare. 

\section{Conclusion}\label{sec:concl}

From the comparison of 2MASS star counts with a model of population synthesis in the outer regions of the Galaxy we investigate the external disc structure and in particular the warp feature followed by stars. We obtain a good fit of one side of the disc with a scale length $h_R$=2200 pc and a simple warp model following a slope of 0.09 and starting at 8.4 kpc. On the other side, at negative longitudes, the fit remains unsatisfactory either with a S shape warp, or a Fourier modes shape. A shorter truncation of the disc improves a bit the fitting at negative longitudes but degrades it at positive longitudes. This could be an indication of a variation of the distance at which the stellar disc stops as a function of azimuth. Despite the uncertainties, it is clear from our study of the 2MASS star counts that the warp is less marked in stars than in the gas and it also shows different shapes of the disc at negative and positive longitudes. There is also a slight tendancy for the dust warp to be intermediate between the HI and the stars.
Further studies are needed to confirm these conclusions, in particular a dynamical approach with kinematical data, helpful in understanding the dynamical evolution of this structure, its origin and its link with the gaseous warp.

\begin{acknowledgements}

This publication makes use of data products from the Two Micron All Sky Survey, which is a joint project of the 
University of Massachusetts and the Infrared Processing and Analysis Center/California Institute of Technology, 
funded by the National Aeronautics and Space Administration and the National Science Foundation.
D.J. Marshall is funded by the National Sciences and Engineering Research Council of Canada through its SRO programme. The CDSClient package was used for the remote querying of the 2MASS dataset.

\end{acknowledgements}

\end{document}